# HIGH-EFFICIENCY INDUCTIVE OUTPUT TUBES USING A 3rd HARMONIC DRIVE ON THE GRID

H.P. Freund,[1,2] R.L. Ives,[1] and W. Sessions[3]
[1]Calabazas Creek Research, San Mateo, CA 94404
[2]University of New Mexico, Albuquerque, New Mexico, 87131
[3]Georgia Tech Research Institute, Atlanta, GA 30318

In this paper, we discuss the use of a 3rd harmonic component to the drive voltage on the grid of an inductive output tube (IOT). High-efficiency inductive output tubes (IOTs) are typically characterized by efficiencies up to 70 – 75%. However, the achievement of efficiencies greater than 80% would substantially reduce the operating costs of next-generation accelerators. In order to achieve this goal, we consider the addition of a 3rd harmonic component to the drive signal on the grid of an inductive output tube. The use of a 3rd harmonic drive component in IOT guns has been considered in order to apply such a gun as the injector [1] of radio frequency linear accelerators (RF linacs). Here we consider that the IOT will be used to provide the rf power to drive RF linacs and apply the 3rd harmonic with the intention of increasing the efficiency of the RF output of the IOT. We consider a model IOT with a 700 MHz resonant cavity and using an annular beam with a voltage of 30 kV, an average current of 6.67 A yielding a perveance of about 1.3 μP. We simulate this IOT using the NEMESIS simulation code [2] which has been successfully validated by comparison with the K5H90W-2 IOT developed by Communications and Power Industries. It is found that the effect of the 3rd harmonic on the efficiency is greatest when the phase of the 3rd harmonic is shifted by $\pi$ radians with respect to the fundamental drive signal and with 3rd harmonic powers greater than about 50% that of the fundamental drive power. For the present example, we show that efficiencies approaching 86% are possible by this means.

## I. INTRODUCTION

High-efficiency inductive output tubes (IOTs) are typically characterized by efficiencies up to 70 – 75%. However, the achievement of efficiencies greater than 80% would substantially reduce the operating costs of next-generation accelerators. In order to achieve this goal, we consider the addition of a 3rd harmonic component to the drive signal on the grid of an inductive output tube. The use of a 3rd harmonic drive component in IOT guns has been considered in order to apply such a gun as the injector [1] of radio frequency linear accelerators (RF linacs). Here we consider that the IOT will be used to provide the drive power for RF linacs and apply the 3rd harmonic with the intention of increasing the efficiency of the RF output of the IOT.

We consider a model IOT with a 700 MHz resonant cavity and using an annular beam with a voltage of 30 kV, an average current of 6.67 A yielding a perveance of about 1.3 μP. We simulate this IOT using the NEMESIS simulation code [2] which has been successfully validated by comparison with the K5H90W-2 IOT developed by Communications and Power Industries. It is found that the effect of the 3rd harmonic on the efficiency is greatest when the phase of the 3rd harmonic is shifted by $\pi$ radians with respect to the fundamental drive signal and with 3rd harmonic powers greater than about 50% that of the fundamental drive power. For the present example, we show that efficiencies approaching 86% are possible by this means.

The organization of the paper is as follows. The numerical model incorporated in the NEMESIS simulation code is described in Section. II. The simulation results for the 700 MHz IOT both with and without the 3rd harmonic drive is described in Section II, and a summary and discussion is given in Section IV.

## II. THE NEMESIS SIMULATION CODE

The numerical formulation implemented in the NEMESIS code [2] contains elements similar to what is employed in particle-in-cell simulation codes. Integration of the dynamical equations is performed in time, so the code can treat implicitly particles that might turn around. This can be important in high-efficiency designs where particles lose a great deal of energy.

NEMESIS contains an equivalent (LRC) circuit model for the cavity voltage with a model for the circuit fields taken from Kosmahl and Branch [3] which is scaled using the cavity voltage. The integration of particle trajectories makes use of the circuit fields obtained in this fashion as well as an analytic model for the focusing fields and a two-dimensional Poisson solver for the space charge fields.

The numerical procedure is illustrated in Fig. 1. The procedure in stepping from $t \rightarrow t + \Delta t$ begins with a 4th order Runge-Kutta integration of the equivalent circuit equations. We typically take 100 steps per wave period to ensure accuracy in this calculation. Once the circuit equations have been stepped, electrons are injected into the cavity if the time coincides with the bunch phase. We inject $N$ charge rings of charge. The charge associated with $i^{th}$ ring is $(2i - 1)I_b(t)\Delta t/N^2$. After injection, the trajectories are integrated using a Boris push [4], which is a 2nd order accurate, two-step process in which the particle momenta are integrated first followed by the particle positions. At this point, the source current is calculated by averaging the current obtained using ($\mathbf{x}_t$, $\mathbf{v}_{t+\Delta t/2}$) and ($\mathbf{x}_{t+\Delta t}$, $\mathbf{v}_{t+\Delta t/2}$). This ensures 2nd order accuracy for the overall procedure. Finally, we test whether any particles have left the system (from either end of the cavity or by striking the wall), and, if necessary, eject them from the simulation. This is repeated as many times as necessary to simulate any given pulse time.

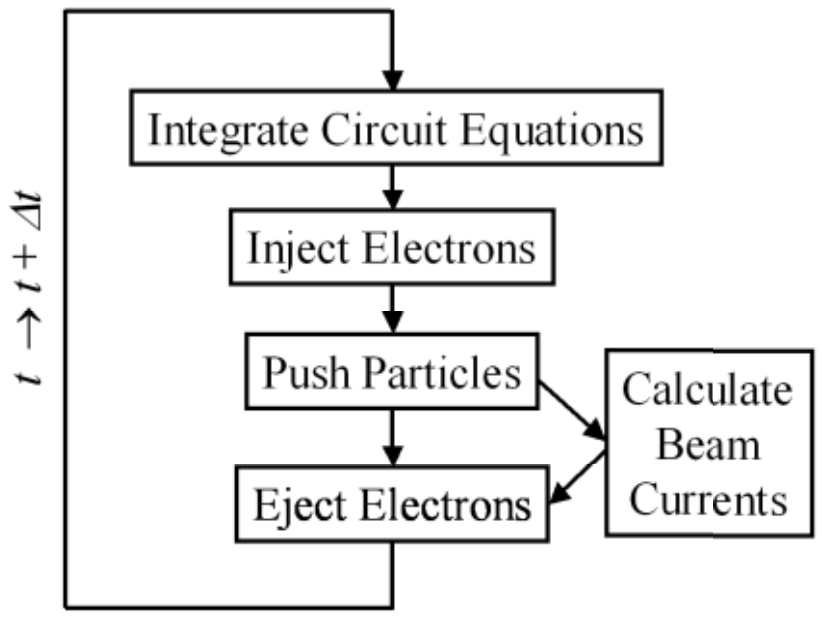

Fig. 1: Flow chart of the simulation procedure [2].

The inclusion of the 3rd harmonic is done via the following model for the drive current, where $I_p$ is the peak current, $\varepsilon$ denotes the ratio of the 3rd harmonic to the fundamental and $\varphi$ is the phase shift of the 3rd harmonic relative to the fundamental.

$$I(t) = I_p \left[ \sin\left( \pi \frac{t}{\tau_{width}} \right) + \varepsilon \sin\left( 3\pi \frac{t}{\tau_{width}} + \varphi \right) \right]^2 , \quad (1)$$

for $t < \tau_{width} < 1/f$ and zero otherwise, where $f$ is the wave frequency and $\tau_{width}$ denotes the portion of the wave period over which electrons are drawn off the grid. The ratio of the average to peak current using this model for the drive current is

$$\frac{I_{avg}}{I_p} = \frac{1+\varepsilon^2}{2} \frac{\tau_{width}}{\tau_{period}} . \quad (2)$$

## III. SIMULATION RESULTS

A schematic illustration of an IOT is shown in Fig. 2 which employs a solid beam. In contrast, here we consider an annular beam. An IOT is composed of three elements: the gridded gun, the output cavity and the collector. Here we simulate the output cavity for a given injected electron beam both with and without the 3rd harmonic drive, where our primary purpose is to investigate the effect of the 3rd harmonic.

The configuration under study consists in a cavity tuned to a resonant frequency of 700 MHz with $R/Q$ = 100 Ω and a loaded $Q$ = 84.6 with a radius of 9.009 cm, a length of 9.144 cm, and where the center of the gap is located 4.571 cm downstream from the entrance to the cavity. The optimal gap length was found to be about 3.1 cm in the absence of the 3rd harmonic. The electron beam voltage and current are 30 kV and 6.67 A respectively, and is injected as an annulus centered about the axis of symmetry with an inner radius of 5.128 cm and an outer radius of 8.079 cm. A solenoidal focusing field is used with an amplitude of 126 G which is close to the Brillouin field. The ratio of the average to peak current is 0.15. In the absence of any 3rd harmonic drive, this implies that the ratio of the width of the pulse to the resonant period is 0.30.

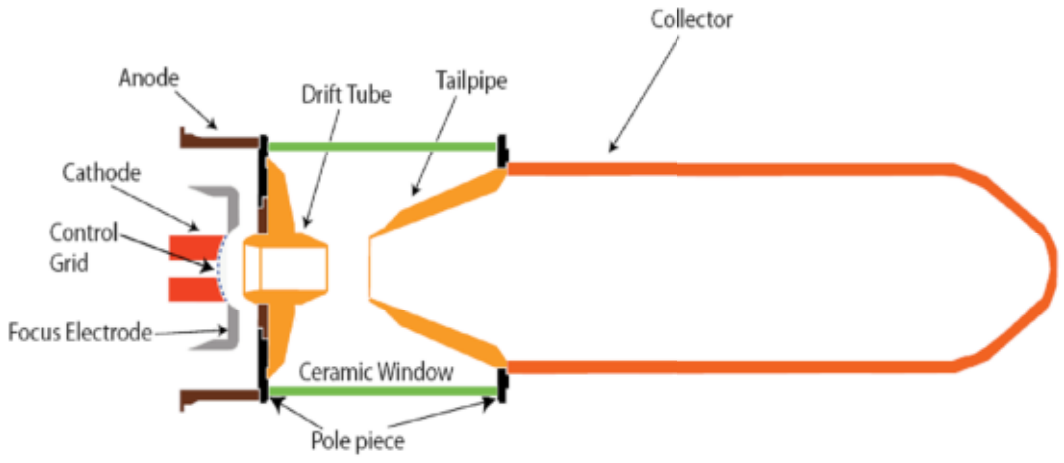

Fig. 2: Schematic illustration of an inductive output tube, where the output cavity is located between the pole pieces [2].

We first consider the performance in the absence of the 3rd harmonic drive. A plot showing the output power as a function of time is shown in Fig. 3 for the above-mentioned parameters. As shown in the figure, the power increases to a steady-state after about 50 – 60 nsec. The plot shows the instantaneous power, which is the product of the voltage and current across the load. Since both the voltage and current vary as cos($\omega t$), the product has a bulk (or average) component as well as a harmonic oscillation that varies as cos($2\omega t$). It is this rapid oscillation that gives the figure the appearance of a solid rising bar. However, the average power is half that of the maximum and reaches a steady-state of about 170 kW. Note that this already corresponds to an efficiency of about 84 – 85 percent. The linewidth of this configuration is 3.1%, as shown in Fig. 4.

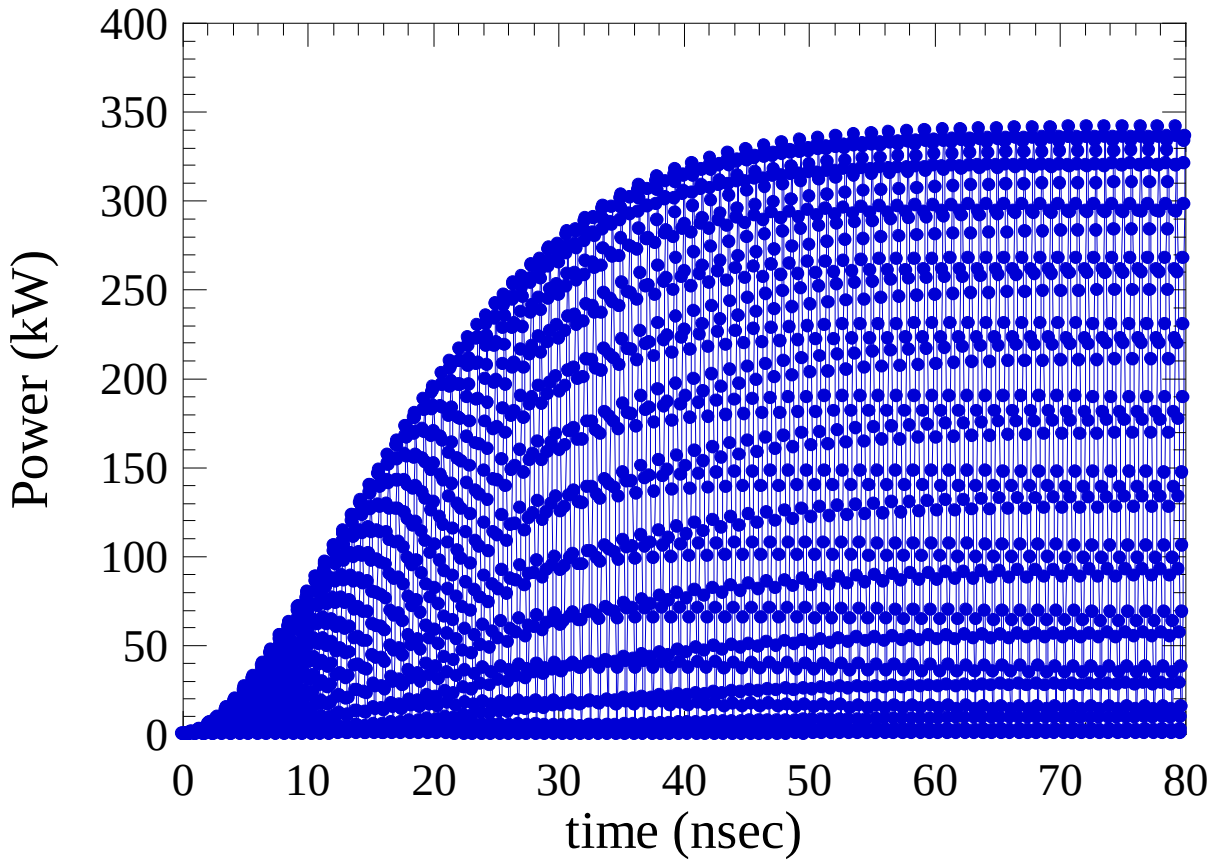

Fig. 3: Output power vs time.

The results shown in Figs. 3 and 4 were obtained using the optimal cavity/beam parameters. In order to illustrate the sensitivity of the design to cavity/beam parameters we ow consider the effect of various parameters on the performance.

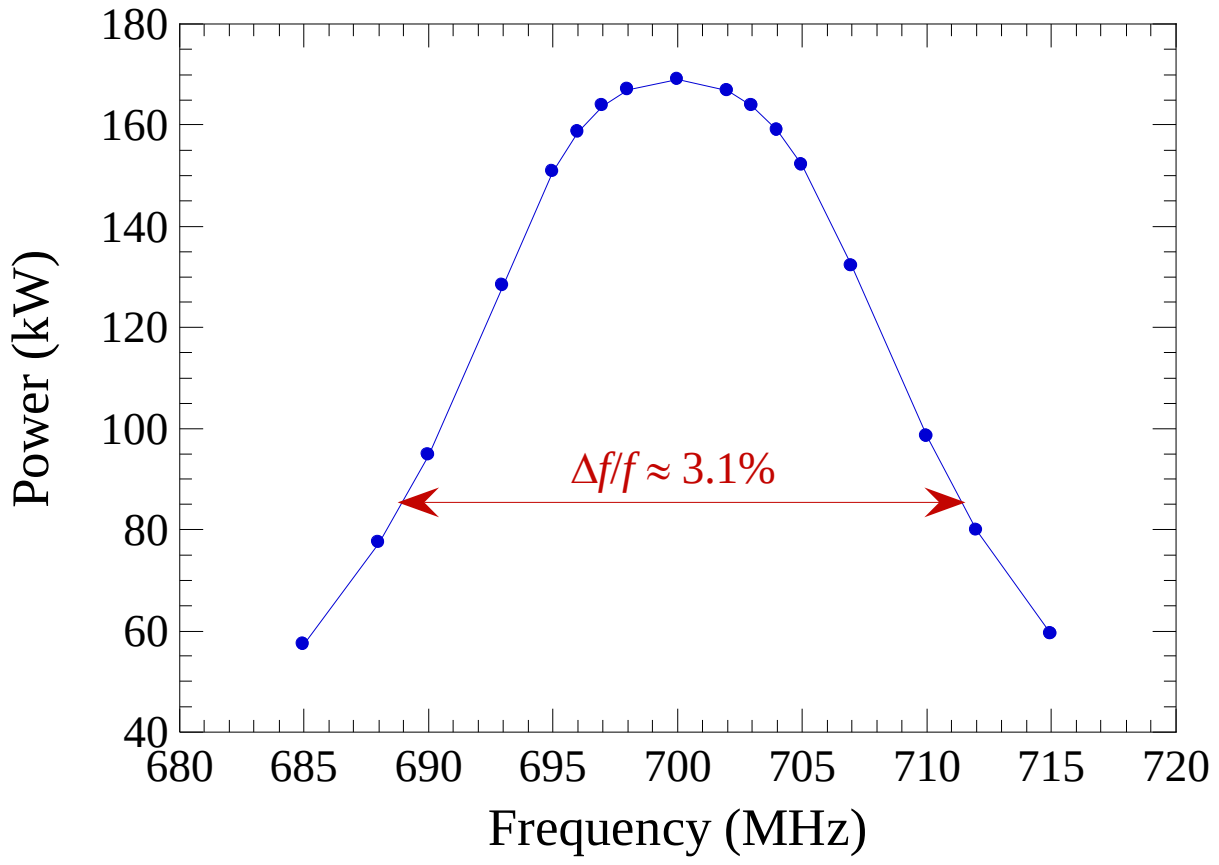

Fig. 4: Plot of the linewidth.

We first consider the sensitivity of the performance to the gap. The performance is relatively insensitive to the position of the gap center, but is more sensitive to the gap length, which is shown in Fig. 5. It is clear from the figure that the optimal gap length is about 3.1 cm.

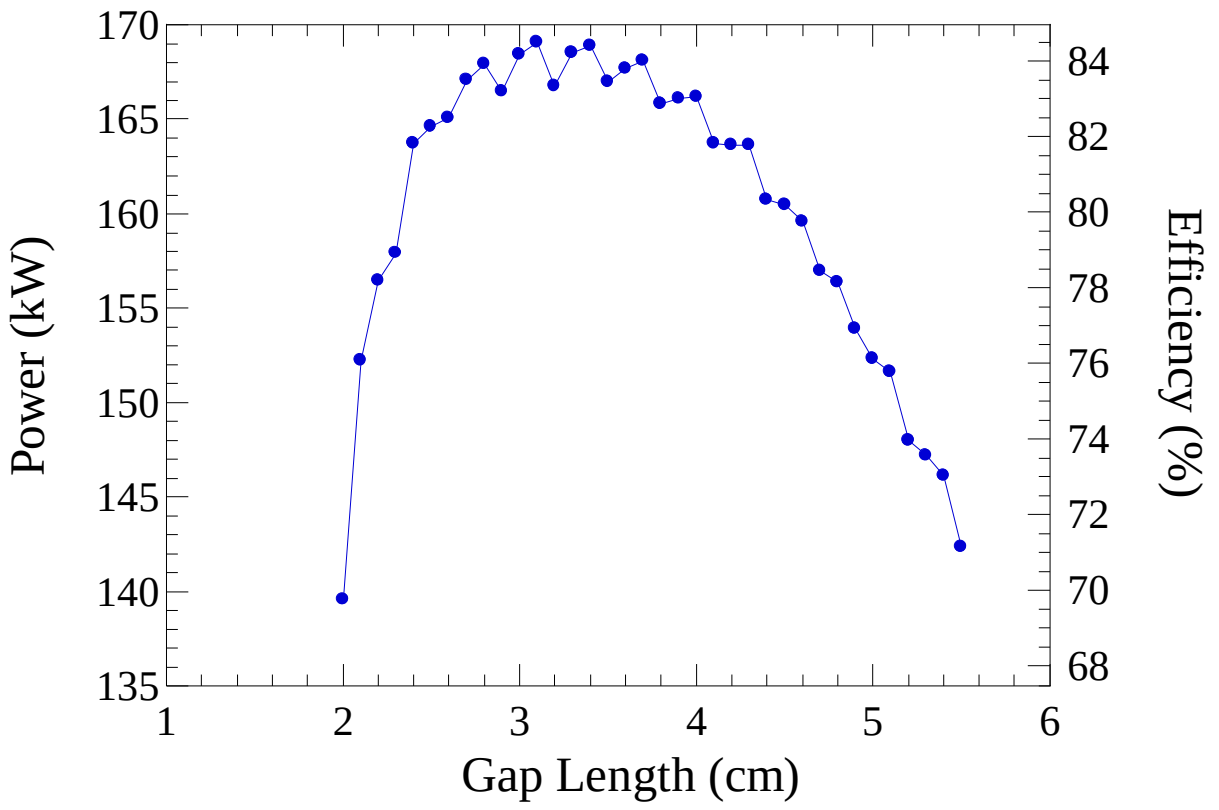

Fig. 5: Sensitivity to the gap length.

The effect of the 3rd harmonic on the performance of the IOT depends on the phase relationship of the harmonic with respect to the fundamental. This is evident in Fig. 6 where we plot the output power (left axis) and the efficiency (right axis) versus the phase of the 3rd harmonic (blue) as well as the performance in the absence of the 3rd harmonic (red). As shown in the figure, the effect of the 3rd harmonic is maximized for a phase shift of about $\pi$ radians. At that point, the efficiency reaches approximately 85.5%. This compares with an efficiency of about 84.5% in the absence of the 3rd harmonic.

The 3rd harmonic results shown in Fig. 6 were obtained using $\varepsilon = 1$. The variation in the effect of the 3rd harmonic with harmonic power is an important consideration in the design of the IOT, and we show the variation in performance with $\varepsilon$ in Fig. 7 for the optimal $\pi$ phase shift. It is clear from the figure that the optimal effect of the 3rd harmonic is found for $\varepsilon$ greater than about 0.5 but is relatively insensitive to increases beyond that value (86.3%).

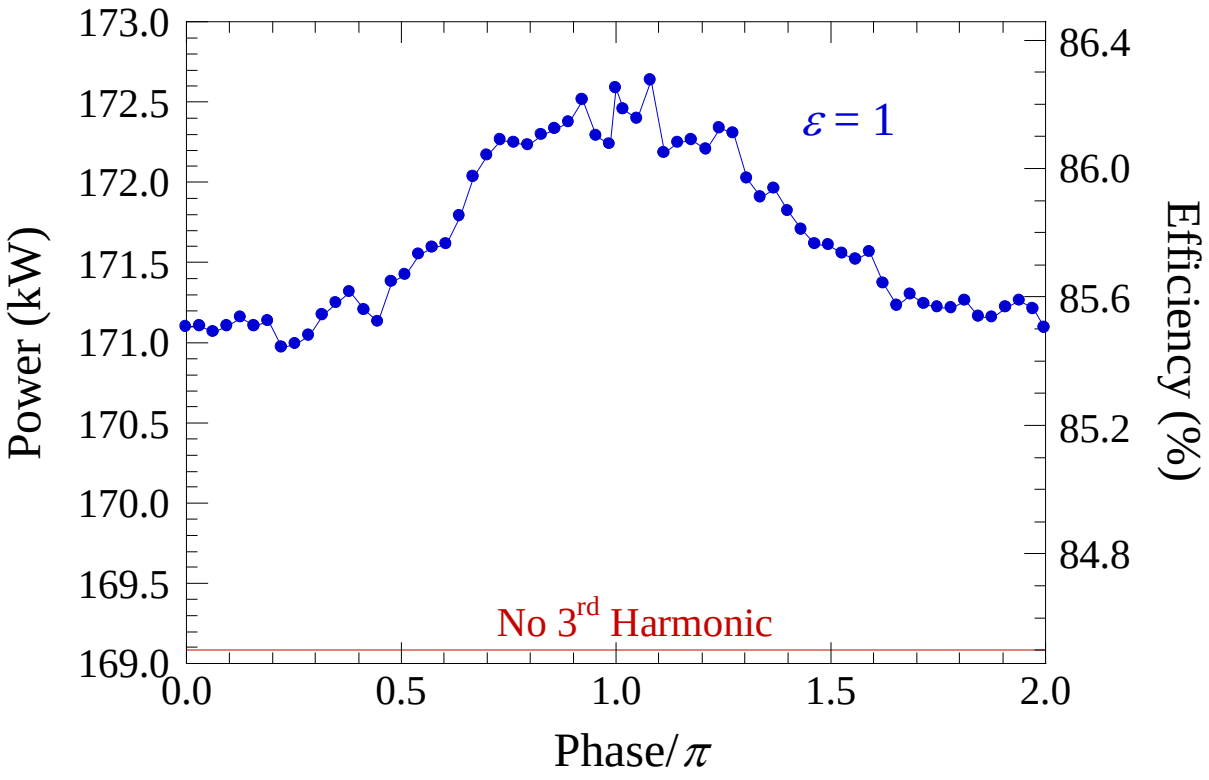

Fig. 6: IOT performance with 3rd harmonic drive on the grid.

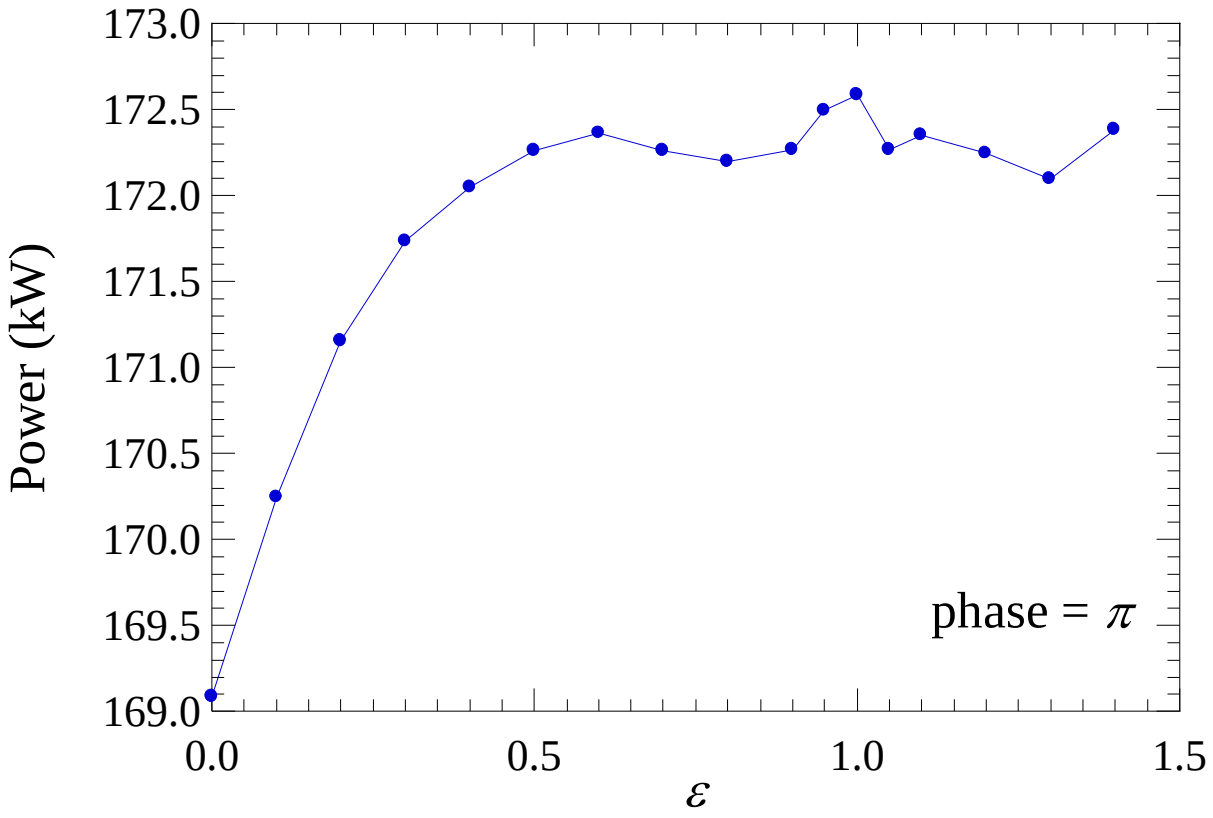

Fig. 7: Variation in performance with the power in the 3rd harmonic drive.

## IV. SUMMARY AND CONCLUSION

In this paper, we have described a model for the inclusion of a 3rd harmonic component on the drive of a high-efficiency IOT. It is found that the effect of the 3rd harmonic on the efficiency is greatest when the phase of the 3rd harmonic is shifted by $\pi$ radians with respect to the fundamental drive signal and with 3rd harmonic powers greater than about 50% that of the fundamental drive power. For the present example, we show that efficiencies approaching 86% are possible by this means. Results in progress for a multi-beam configuration show that improvements in the performance due to the 3rd harmonic can reach 5 – 10%, and this will be reported in a future paper.

## ACKNOWLEDGEMENTS

Work supported by the US Department of Energy under Award No. DE-SC0019800. The authors acknowledge the assistance of engineers at Communications & Power Industries, Inc.